\documentclass[10pt,conference]{IEEEtran}
\IEEEoverridecommandlockouts

\usepackage{cite}
\usepackage{amsmath,amssymb,amsfonts}
\usepackage{algorithmic}
\usepackage{graphicx}
\usepackage{textcomp}
\usepackage{xcolor}
\usepackage{censor}

\usepackage{url}
\usepackage{array}
\usepackage{fvextra}
\usepackage{listings}
\usepackage{float}

\definecolor{cloudwhite}{cmyk}{0,0,0,0.025}
\usepackage{color} 
\definecolor{mygreen}{rgb}{0,0.6,0}
\definecolor{mygray}{rgb}{0.5,0.5,0.5}
\definecolor{mymauve}{rgb}{0.58,0,0.82}
\definecolor{myblue}{rgb}{0.3,0.3,1}
\definecolor{myorange}{rgb}{0.97,0.5,0.16}
\definecolor{darkgray}{rgb}{.4,.4,.4}
\definecolor{purple}{rgb}{0.65, 0.12, 0.82}

\lstdefinelanguage{JavaScript}{
keywords={typeof, new, true, false, catch, function, return, null, catch, switch, var, if, in, while, do, else, case, break, const},
keywordstyle=\color{darkgray}\bfseries,
ndkeywords={class, export, boolean, throw, implements, import, this},
ndkeywordstyle=\color{darkgray}\bfseries,
identifierstyle=\color{black},
sensitive=false,
comment=[l]{//},
morecomment=[s]{/*}{*/},
commentstyle=\color{mygreen}\ttfamily,
stringstyle=\color{blue}\ttfamily,
morestring=[b]',
morestring=[b]"
}
 
\lstdefinelanguage{Gherkin}{
keywords={Feature, Scenario, Given, When, Then, And, As a, I want, So that},
keywordstyle=\color{darkgray}\bfseries,
basicstyle=\footnotesize\ttfamily,
showstringspaces=false,
numbers=none
}

\lstdefinelanguage{US}{
keywords={As, a, I, want, So, that},
keywordstyle=\color{darkgray}\bfseries,
basicstyle=\footnotesize\ttfamily,
showstringspaces=false,
numbers=none
}

\def\BibTeX{{\rm B\kern-.05em{\sc i\kern-.025em b}\kern-.08em
    T\kern-.1667em\lower.7ex\hbox{E}\kern-.125emX}}
\begin{document}

\title{Acceptance Test Generation with Large Language Models: An Industrial Case Study}

\author{\IEEEauthorblockN{Margarida Ferreira}
\IEEEauthorblockA{\textit{Critical TechWorks}\\
\textit{and Faculty of Engineering}, \\ \textit{University of Porto} \\
Porto, Portugal\\
up201905046@up.pt}
\and
\IEEEauthorblockN{Luís Viegas}
\IEEEauthorblockA{\textit{Critical TechWorks}\\
\textit{and Faculty of Engineering}, \\
\textit{University of Porto} \\
Porto, Portugal\\
up201904979@up.pt}
\and
\IEEEauthorblockN{João Pascoal Faria}
\IEEEauthorblockA{\textit{INESC TEC,}\\
\textit{Faculty of Engineering,}\\ 
\textit{University of Porto} \\
Porto, Portugal\\
jpf@fe.up.pt}
\and
\IEEEauthorblockN{Bruno Lima}
\IEEEauthorblockA{\textit{LIACC,}\\
\textit{Faculty of Engineering,}\\ 
\textit{University of Porto} \\
Porto, Portugal\\
brunolima@fe.up.pt}
}

\maketitle

\begin{abstract}
Large language model (LLM)-powered assistants are increasingly used for generating program code and unit tests, but their application in acceptance testing remains underexplored. 
To help address this gap, this paper explores the use of LLMs for generating executable acceptance tests for web applications through a two-step process: (i) generating acceptance test scenarios in natural language (in Gherkin) from user stories, and (ii) converting these scenarios into executable test scripts (in Cypress), knowing the HTML code of the pages under test.
This two-step approach supports acceptance test-driven development, enhances tester control, and improves test quality.
The two steps were implemented in the AutoUAT and Test Flow tools, respectively, powered by GPT-4 Turbo, and integrated into a partner company’s workflow and evaluated on real-world projects.
The users found the acceptance test scenarios generated by AutoUAT helpful 95\% of the time, even revealing previously overlooked cases. Regarding Test Flow, 92\% of the acceptance test cases generated by Test Flow were considered helpful: 60\% were usable as generated, 8\% required minor fixes, and 24\% needed to be regenerated with additional inputs; the remaining 8\% were discarded due to major issues.
These results suggest that LLMs can, in fact, help improve the acceptance test process, with appropriate tooling and supervision.

\end{abstract}


\section{Introduction}\label{sec:intro}

Recent advances in generative artificial intelligence (AI) and large language models (LLMs) \cite{brown2020language} have enabled AI-powered assistants, like GitHub Copilot\footnote{\url{https://github.com/features/copilot}}, capable of accurately generating code and unit tests from textual prompts~\cite{kn:mpaolo2023}, enhancing developer and tester productivity~\cite{wermelinger2023using, peng2023impact, blueoptima2024impact}. 
However, LLMs are still underexplored for acceptance test (AT) generation \cite{wang2024software}—the highest test level, focused on validating that the software meets user expectations and business requirements~\cite{kn:atcontext}.

To help address this gap, this paper explores the use of LLMs to improve the acceptance test process, especially in the context of agile processes and practices, such as: (i) capturing user requirements as user stories (US); (ii) refining them with (user) acceptance tests written in a restricted natural language (NL) accessible for non-technical users and amenable for subsequent automation, like Gherkin\footnote{\url{https://cucumber.io/docs/gherkin/}} \cite{hellesoy2017cucumber}, following    
acceptance test-driven development (ATDD) \cite{gartner2013atdd} and behaviour-driven development (BDD) \cite{kn:aat-2} principles; (iii) automating them using popular frameworks like Selenium\footnote{\url{https://www.selenium.dev/}} or Cypress\footnote{\url{https://www.cypress.io/}} \cite{garcia2024exploring}, to reduce manual testing~\cite{kn:aat} and enable continuous integration (CI). 

More specifically, we aim to partially automate the last two activities by leveraging LLMs’ natural language understanding and code generation capabilities to: (i) automatically generate acceptance test scenarios from user stories, and (ii) automatically convert these scenarios into executable test scripts.

Our research addresses two key questions: 
\begin{itemize}
    \item \textbf{RQ1}: To what extent can LLMs help in generating accurate and comprehensive acceptance test scenarios in natural language from user stories?
    \item \textbf{RQ2}: To what extent can LLMs help in generating accurate and complete\footnote{By 'complete', we mean covering all steps.} executable test scripts from acceptance test scenarios in natural language?
\end{itemize}

To address them, we implemented the two generation steps in separate tools—AutoUAT and Test Flow—which can be used independently or together. To ease integration and evaluation in a partner company in the automotive sector, we used GPT-4 Turbo as the LLM, Gherkin and TypeScript as the target languages, and Cypress as the test automation framework.

While assessing long-term productivity and quality gains would require further studies, we expect to help address common industry issues: (i) AT writing is often seen as tedious and low-priority, risking missed edge cases; and (ii) test script writing adds workload for developers, potentially shifting focus from understanding requirements to test success~\cite{kn:aat}.

The main contributions of our work are:

\begin{itemize}
    \itemsep 0em
    \item \textbf{AutoUAT} - an LLM-powered tool, integrated in an industrial environment, that automatically generates acceptance test scenarios in Gherkin from user stories;
    \item \textbf{Test Flow} - an LLM-powered tool, integrated in an industrial environment, that automatically generates executable test scripts for end-to-end web application testing in Cypress, based on Gherkin test scenarios' specifications and the HTML code of the web pages under test purged from irrelevant information, with minimal user intervention; 
    \item \textbf{Evaluation results} - results of evaluation studies in our automotive partner (Critical TechWorks), with 95\% of acceptance test scenarios generated by AutoUAT and 92\% of acceptance test scripts generated by Test Flow considered helpful, and very positive user feedback.
    \end{itemize}

The rest of the paper is organized as follows:
Sec. \ref{sec:approach} describes our test generation approach. Sec. \ref{sec:integration} outlines the tool implementation and integration in an industrial workflow.
Sec. \ref{sec:validation} presents experimental results and user feedback, addressing our research questions. Sec. \ref{sec:related} discusses related work, and Sec. \ref{sec:concl} concludes with key findings and future directions.

\section{Acceptance Test Generation Approach}\label{sec:approach}

\subsection{Overall Approach}\label{ss:overall}

\begin{figure}[h]
    \centering 
    \includegraphics[width=\linewidth]{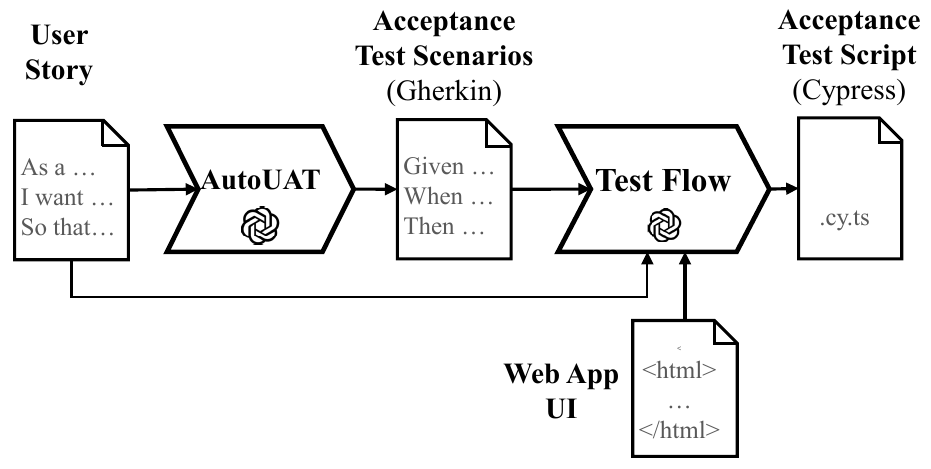} 
    \caption{Acceptance Test Generation with AutoUAT and Test Flow.} 
    \label{fig:approach} 
\end{figure}

Our two-step approach for generating executable acceptance test scripts from user stories is outlined in Fig. \ref{fig:approach}.

The first LLM-powered tool — AutoUAT — was designed to generate Gherkin acceptance test scenarios directly from user story titles and descriptions.
The second LLM-powered tool — TestFlow — then takes these Gherkin test scenarios as input, along with the initial user story description and the HTML code of the web pages under test, to generate executable test scripts in Cypress (in TypeScript syntax).

This approach was designed to effectively support ATDD:
NL test scenarios can be generated when new user stories are created to clarify requirements and guide implementation; once the application user interface (UI) is ready, executable test scripts can be created, mapping test steps in the test scenarios to specific actions and UI elements.
It also gives testers and other stakeholders more control, as they can review the generated test scenarios and create their own before converting them to executable test scripts, improving test quality.

Model selection and prompt engineering are described in the next subsections. 

\subsection{Model Selection}


The study was carried out in an industrial environment, where privacy and security were primary concerns. To address these, the solution was to use a model hosted within our industrial partner's Azure environment. We evaluated several cutting-edge models available in Azure, including OpenAI's GPT models and Meta's CodeLlama family.


Among the CodeLlama family, only the CodeLlama Instruct 70B model had benchmarks comparable to the GPT family, but it required a high-cost Azure machine with multiple GPUs to run. In contrast, Azure offers a pay-as-you-go service called Azure OpenAI, which, as of February 2024, includes the latest GPT-4 Turbo model with a 128,000 token context window\footnote{\url{https://platform.openai.com/docs/models/gpt-4-turbo-and-gpt-4}}. This model outperformed CodeLlama Instruct 70B in benchmarks like HumanEval \cite{chen2021evaluating} and MBPP \cite{austin2021program}. Furthermore, the 'gpt-4-1106-preview' version was available at a promotional price, making it the preferred choice due to its superior performance, cost-effectiveness, and integration within the industrial partner's Azure environment.

\subsection{From User Stories to Acceptance Test Scenarios}\label{ss:story_to_cases}

This phase aims to derive acceptance tests from user stories, translating high-level requirements into detailed Gherkin scenarios to validate system behavior, ensure shared understanding among stakeholders, and guide implementation.

The inputs for this phase include: 
\begin{itemize} 
    \item \textbf{User Story Title:} A concise, descriptive name capturing the feature or functionality from the user’s perspective.
    \item \textbf{User Story Description:} A detailed outline of functionality, including requirements, acceptance criteria, and expected outcomes. 
\end{itemize}

The output is a set of acceptance test scenarios in Gherkin format, covering key aspects of the user stories for comprehensive validation.

To achieve accurate, comprehensive, and robust outputs across diverse scenarios, an iterative prompt engineering process was applied to optimize clarity, context awareness, and response specificity --- see the final prompt in Appendix \ref{sec:AutoUATprompt}.

\subsection{From Acceptance Test Scenarios to Executable Test Scripts}\label{ss:cases_to_script}

The inputs for this phase are the user story to be tested, the corresponding Gherkin acceptance test scenarios, and the HTML code of the pages under test. The output is a Cypress test script automating all the specified Gherkin scenarios.


Initial experiments indicated that fine-tuning was unnecessary, as the model already demonstrated a solid understanding of Cypress and TypeScript syntax. This was expected due to the widespread adoption of these technologies. Instead, the focus shifted to prompt engineering, involving experimental adjustments to the prompt in order to provide the model with the necessary context for generating accurate test scripts.

The following metrics were used to assess model's performance and identify areas for improvement during the prompt engineering process:

\begin{itemize}
    \itemsep 0em
    \item \textbf{Syntactic Correctness:} The proportion of generated tests adhering to TypeScript syntax.
    \item \textbf{Semantic Relevance:} The proportion of generated tests reflecting the inputted Gherkin scenarios.
    \item \textbf{Accssibility:} The proportion of generated tests with comprehensive natural language comments.
    \item \textbf{Prompt Size:} An evaluation of both input prompt and response sizes, for cost considerations.
\end{itemize}

The prompt engineering process involved the following steps: gathering a set of inputs, constructing a basic prompt, 
evaluating the performance metrics against the inputs, and iteratively refining the prompt to optimize the metrics that demonstrated weaker performance, following an experimental approach.

The final prompt consists of a user and a system prompt. 

The user prompt contains information specific to the feature under test (to be collected automatically), including:

\begin{figure*}
    \centering 
    \includegraphics[width=\textwidth]{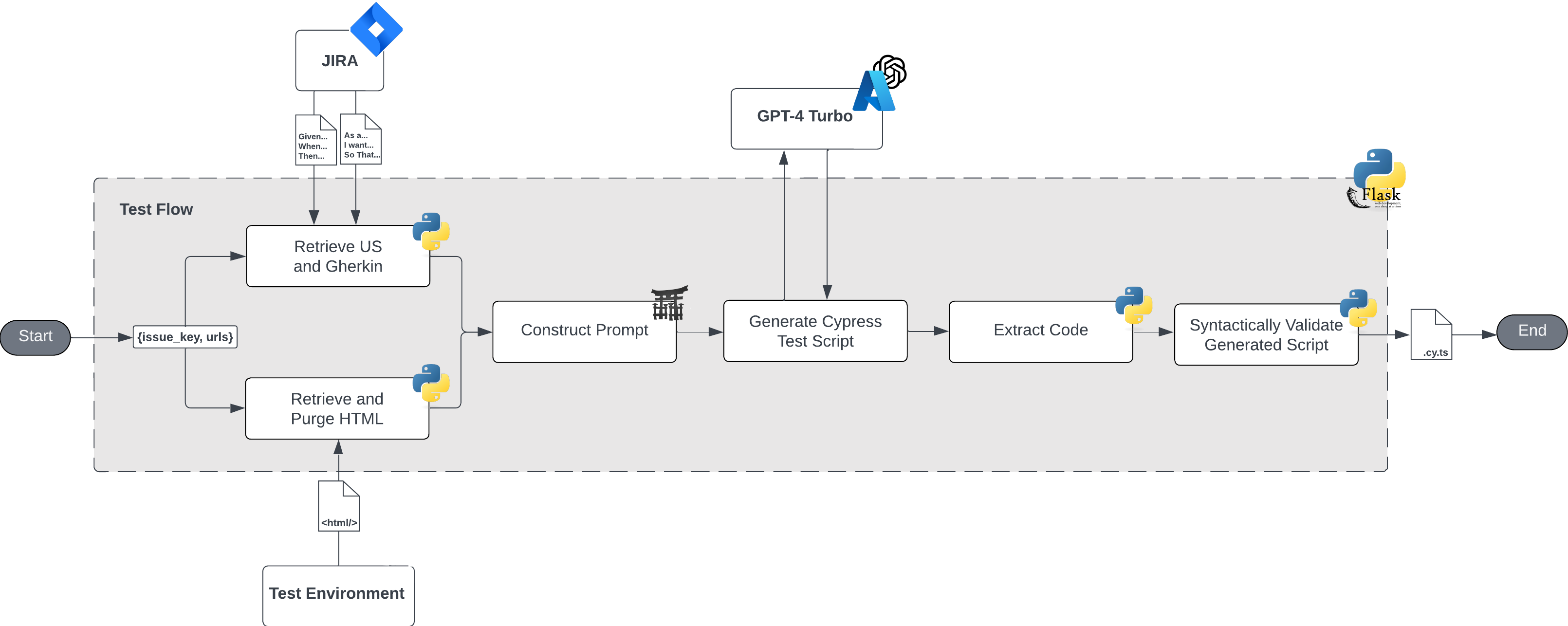} 
    \caption{Flow Diagram of Test Flow.} 
    \label{fig:test-flow} 
\end{figure*}

\begin{figure*}
    \centering 
    \includegraphics[width=\textwidth]{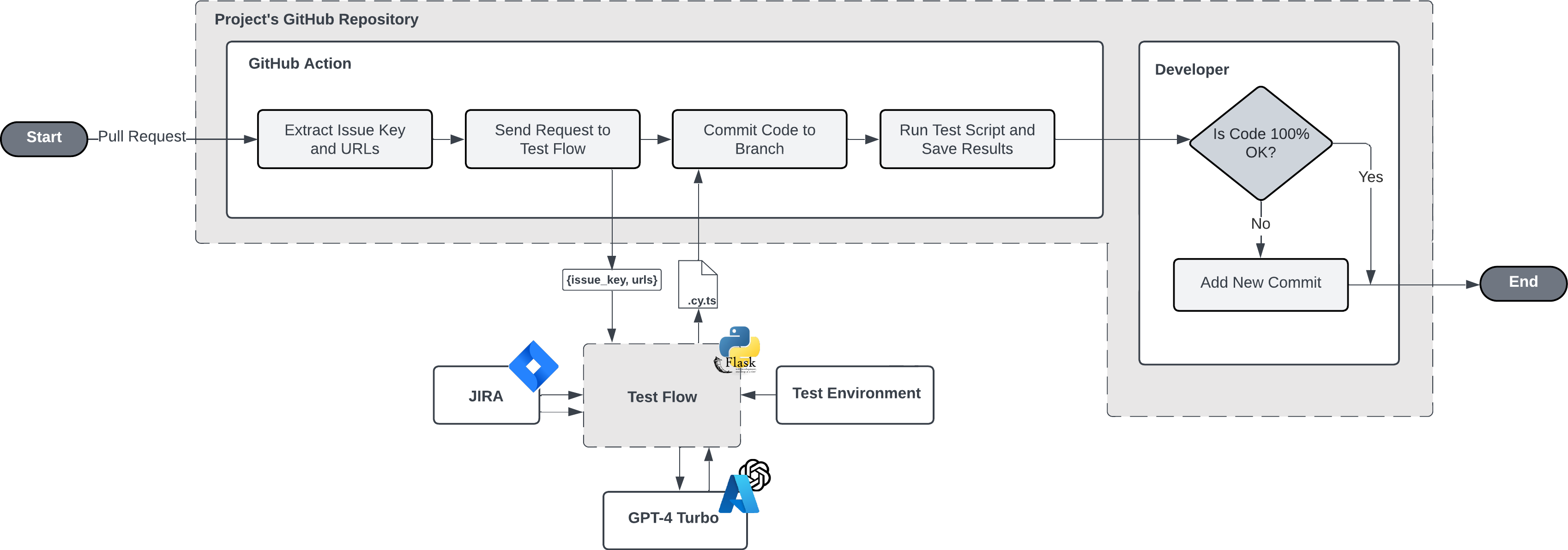} 
    \caption{GitHub Action for Integrating Test Flow into Current Workflows.} 
    \label{fig:action} 
\end{figure*}

\begin{itemize}
    \itemsep 0em
    \item \textbf{User Story:} For context on the feature being tested.
    \item \textbf{Gherkin Scenarios:} For context on the test cases to generate.
    \item \textbf{Pages' HTML:} For context on the pages' elements and their identifiers.
\end{itemize}

To reduce costs related to the prompt size, the HTML is preprocessed to remove style and script elements, as these were found unnecessary for the model's understanding of the HTML, and did not compromise the performance metrics. 

The system prompt provides additional context on the product and guides the model's response, including (see Appendix  \ref{sec:TestFlowPrompt}):
\begin{itemize}
    \itemsep 0em
    \item \textbf{Product Context:} For additional context on the product being tested.
    \item \textbf{Cypress Good Practices:} For adherence to Cypress best practices.
    \item \textbf{Other Aspects:} For addressing specific situations the model was overlooking.
\end{itemize}

\section{Integration in the Industrial Environment}\label{sec:integration}

This section describes the tools developed to seamlessly incorporate our approach into our industrial partner's operational framework, enabling its evaluation in a real-world context, whilst respecting confidentiality.

\subsection{AutoUAT: From User Stories to Acceptance Test Scenarios}\label{ss:story_to_cases_integration}  

To integrate the process of transforming user stories into acceptance test scenarios in an industrial environment and gather user feedback and usage information, an automated workflow was developed using the Microsoft Power Platform. 

The process begins with users filling out a form with the title and description of the user story. 
Its submission triggers an HTTP request to an Azure container where the LLM is hosted. The LLM processes the prompt provided by the tool with user story information and system instructions and generates the corresponding ATs.
The generated ATs are then automatically sent back to the user via a private message on Microsoft Teams, who can also fill out a feedback form.

\subsection{Test Flow: From Acceptance Test Scenarios to Test Scripts}\label{ss:cases_to_script_integration}

Test Flow was developed for seamless integration into our partner's workflow with the architecture depicted in Fig. \ref{fig:test-flow}.

To minimize user input and improve usability, Python modules were developed to automatically extract user stories and Gherkin test scenarios from JIRA\footnote{\url{https://www.atlassian.com/software/jira}}, given the issue key, and to retrieve HTML code of the pages under test stripped from irrelevant style and script elements, given their URLs.
Then, the prompt is constructed following prompt engineering results and sent to GPT-4 Turbo. Upon receiving the model's response, the generated code is extracted, validated for TypeScript syntax, and finally returned to the user.

Test Flow can be easily integrated with current workflows as it can be invoked through a Flask REST API. As an example of this integration, it was implemented a GitHub Action, illustrated in Fig. \ref{fig:action}. Triggered on a new pull request (PR), it extracts the necessary inputs (issue key and URLs) from the PR description, sends a request with these inputs to Test Flow, commits the generated tests to the source branch of the PR, and executes them. 
As LLM outputs cannot be blindly trusted, developers must review the code generated by Test Flow and implement fixes, when necessary.

\section{Evaluation}\label{sec:validation}

This section details the methods and results of the evaluation efforts conducted with our industrial partner, including experimental evaluations and user surveys. Additionally, it addresses the research questions posed at the beginning and discusses potential threats to the validity of our study. For confidentiality reasons, specific datasets used in this study are not disclosed.

\subsection{AutoUAT Workshop}

 To gather initial user feedback about AutoUAT, we conducted a workshop with six experienced product owners who regularly write user stories and ATs. The session began with a 15-minute introduction to AutoUAT, followed by 15 minutes of hands-on use. Feedback was collected through a survey with 10 questions and an open comments section.

 All participants had over three years of experience and were routinely involved in writing user stories (1 to 10 per sprint) and associated acceptance tests, ensuring feedback relevance.

\begin{figure}
    \centering 
    \includegraphics[scale=0.58]{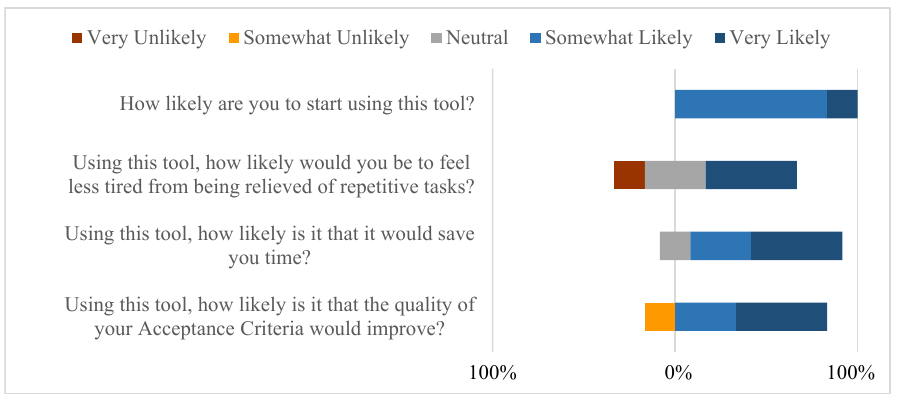} 
    \caption{Workshop Participants' Feedback on AutoUAT Adoption and Benefits.} 
    \label{fig:questions78910} 
\end{figure}

Participants rated the quality of the AutoUAT outputs with an average high score of 8 out of 10.
The likelihood of adoption was unanimously high, likely saving time and improving the acceptance criteria quality (Fig. \ref{fig:questions78910}).


Follow-up comments from the participants further highlight the tool's impact, e.g., \textit{"The tool is extremely helpful. I started writing user stories in Gherkin format, and this tool helped me identify missing acceptance criteria and refine design and UX. It saves me time, and I plan to continue using it."}

\subsection{AutoUAT Usage}

In a second validation phase, the developed tool was made available to everyone within the company for a period of two months, and data was collected on its usage. 

In total, the tool was utilized 166 times by the company's professionals. After each use, users were prompted to provide feedback on the tool outputs. A total of 65 responses were collected. To the question "Was this AT useful?", only 3 respondents answered negatively, while 62 respondents answered positively, representing 95\% of the respondents. 



Additionally, users were encouraged to leave additional comments to complement their responses. Some noteworthy comments include:
\textit{"In this situation, the tool allowed me to think about an error scenario that we were not considering. This had an impact on the product, allowing us to create an error screen in the UX."}

In this case, the user requested the generation of acceptance test scenarios for the "Alphabet User Sign-Up" story described in Appendix \ref{sec:example}. Although the title lacks clarity, the model accurately captured the intended functionality (see Appendix \ref{sec:uatexample}), demonstrating its flexibility in extracting functionality from context and tolerating minor errors. The last two scenarios demonstrate the model's ability to extend beyond the explicit requirements of the user story. 

\subsection{Test Flow Experimental Evaluation}

The validation process for Test Flow started with a quantitative evaluation of the tool for the performance metrics defined in Section \ref{ss:cases_to_script}, using selected issues from a company's e-commerce product. 

In order to ensure data quality and relevance, JIRA issues were selected based on the following criteria:

\begin{itemize}
    \itemsep -0.1em
    \item \textbf{Story Classification:} The issue must be classified as a Story.
    \item \textbf{User Perspective:} The story must be written from the user's perspective.
    \item \textbf{Gherkin Tests:} The issue must have acceptance tests in Gherkin (handwritten or generated with AutoUAT).
    \item \textbf{Quality Assessment:} The issue must have medium to high quality.\footnote{This was subjectively determined based on whether the US provided sufficient context, was unambiguous, and maintained consistent naming conventions.}
    \item \textbf{Team Assignment:} The issue must be preferably assigned to the teams we were most in contact with.\footnote{To facilitate the resolution of potential queries and the validation of outputs. }
\end{itemize}

Given the product's maturity and its current focus on UI modifications and defect resolution rather than new feature development, the sample size was inevitably limited. Nevertheless, we managed to compile a final dataset of 13 issues, each with their US description and Gherkin test scenarios, and gathered the corresponding page URLs for each of them. 

In order to validate the product flow comprehensively, we selected issues that targeted a diverse set of pages. Table \ref{table:story_distribution} illustrates the distribution of the selected issues and Gherkin scenarios according to the related page.

\begin{table}
    \caption{Input subjects for the Test Flow experimental evaluation.}
    \label{table:story_distribution}
    \centering
    \begin{tabular}{|l|c|c|}
        \hline
        \textbf{Web Page}       & \textbf{\#Stories} & \textbf{\#Gherkin Scenarios} \\ \hline
        Product List Page   & 2                         & 9                                    \\ \hline
        Product Detail Page & 3                         & 10                                   \\ \hline
        Cart Page           & 2                         & 7                                    \\ \hline
        Checkout Page       & 4                         & 16                                   \\ \hline
        Orders Page         & 2                         & 8                                    \\ \hline
        \textbf{Total}      & \textbf{13}               & \textbf{50}                          \\ \hline
    \end{tabular}
\end{table}

\begin{figure}
    \centering 
    \includegraphics[width=\linewidth]{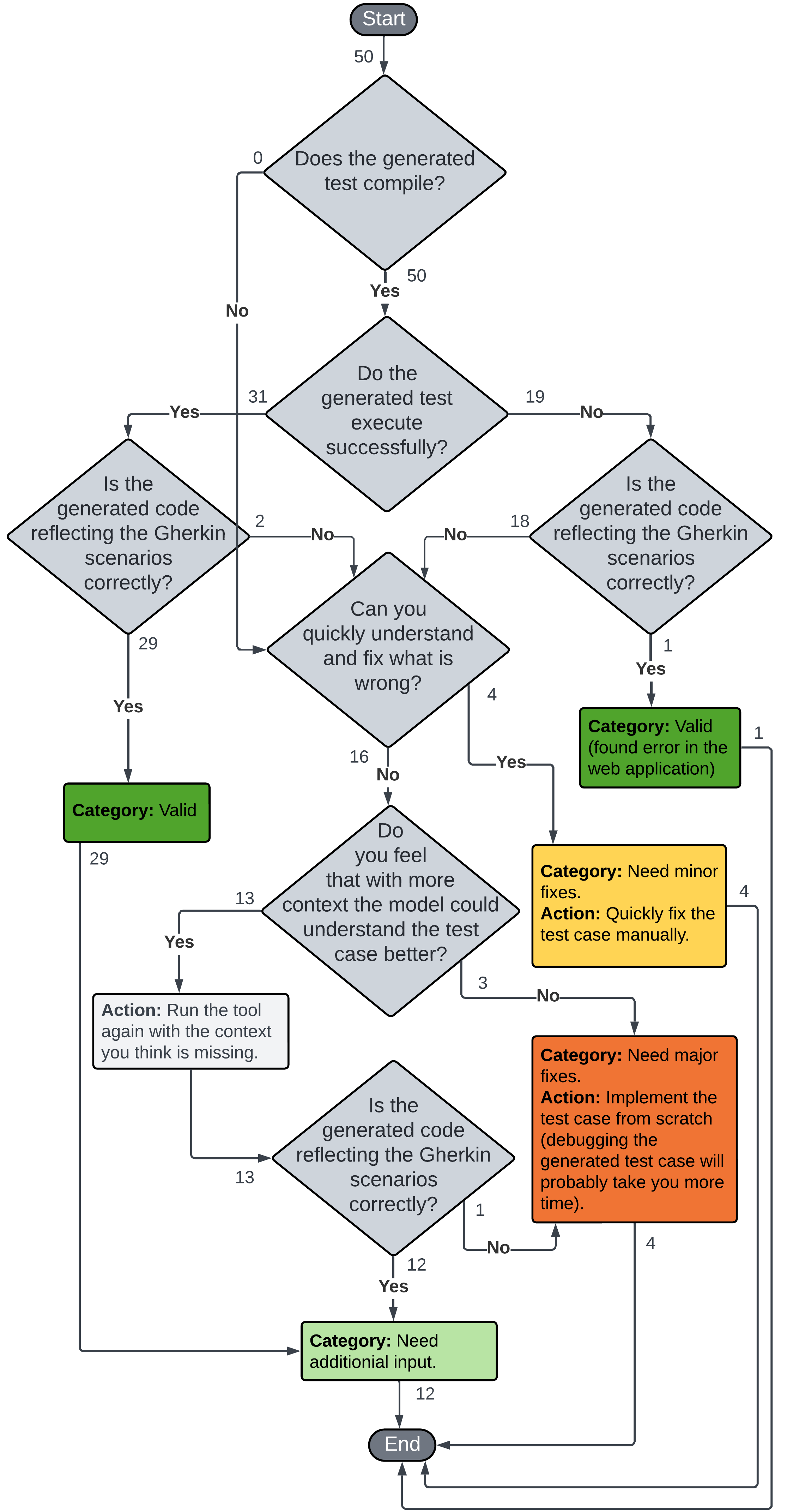} 
    \caption{Procedure for evaluating and classifying the test cases generated with Test Flow. The arrows are labelled with the number of test cases that adhered to the corresponding path in our experiment.} 
    \label{fig:framework} 
\end{figure}

As output, Test Flow generated a Cypress test script for each issue (user story), resulting in 13 scripts and 50 test cases (one per Gherkin scenario) in total. The 50 test cases generated were rigorously evaluated, through a detailed manual analysis, against the previously defined performance metrics. Concurrently, strategies were devised and implemented to rectify the erroneously generated test cases, providing insights into the effort required to correct these test cases. The workflow for this framework and respective results is depicted in Fig. \ref{fig:framework}. The results for the performance metrics are as follows:

\begin{itemize}
    \itemsep -0.1em
    \item \textbf{Syntactic Correctness:} 100\% of the generated test cases adhered to TypeScript syntax.
    \item \textbf{Semantic Relevance:} 60\% of the generated test cases accurately mirrored the original Gherkin scenarios.
    \item \textbf{Accessibility:} 100\% of the generated test cases had comprehensive NL comments.
    \item \textbf{Prompt Size: } Input prompts averaged 9500 tokens; output prompts averaged 750 tokens.
\end{itemize}

The tests generated were syntactically correct and included detailed NL comments, demonstrating the solutions's robustness and accessibility. 
However, the evaluation revealed improvement areas, particularly in semantic relevance, as 40\% of test cases did not accurately reflect the Gherkin scenarios. Manual analysis categorized these test cases into three types:

\begin{itemize}
    \itemsep 0em
    \item \textbf{Lack of Context} (12 cases): Inputted US and Gherkin scenarios lack crucial context for the model to correctly understand the functionality to test.
    \item \textbf{Minor Errors} (4 cases): Errors of less than one line of code that require less than two minutes per test case to comprehend and rectify.
    \item \textbf{Complex Errors}: (4 cases) Remaining errors. In these cases, it is considered not worth the effort to fix them.
\end{itemize}

\begin{figure}
    \centering 
    \includegraphics[scale=0.6]{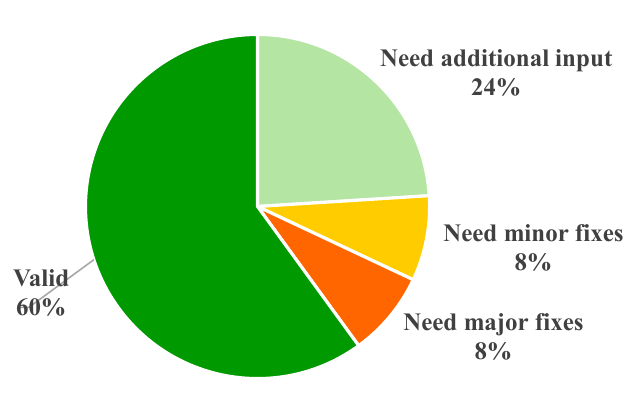} 
    \caption{Classification of 50 Executable Test Cases Generated by Test Flow.} 
    \label{fig:pie_chart} 
\end{figure}

Fig. \ref{fig:pie_chart} provides a visual representation of this distribution. The tests within the first two categories constitute 32\% of the total generated test cases. As depicted in Fig. \ref{fig:framework}, these can be rectified by enriching the input context and implementing minor fixes (one line of code), respectively. Implementing these strategies allowed us to elevate the semantic relevance to 92\%, leaving only the tests with complex errors. In such cases, manual implementation from scratch is recommended.    


Furthermore, a particular failed test revealed a discrepancy; the US and Gherkin tests suggested that a button should be disabled, whereas it was actually hidden. This discrepancy pointed to a flaw in the application being tested, demonstrating Test Flow's defect detection capability.

The size of input and output prompts influences the cost of using Test Flow. Based on the pricing of Azure OpenAI's GPT-4 Turbo as of June 2024, the average cost for generating a test script per user story was 0.12€ in this experiment.

Our investigation found no correlation between specific pages and poorer test outcomes. Instead, test quality was more closely linked to the clarity and context provided in the user stories and Gherkin scenarios.

\subsection{Test Flow Workshop and User Survey}

Following the experimental evaluation of Test Flow, a workshop was conducted with the company’s developers. It included a 15-minute presentation on Test Flow's capabilities, a 15-minute demo of inputs and outputs, and a brief 3-minute survey, which received 16 responses.

The survey targeted individuals who typically conduct Cypress testing, including developers (responsible for creating tests for the features they develop) and Scrum Masters (who also function as developers).
Respondents could optionally disclose their identity for follow-up; 11 did so.

Participant demographics are shown in Table \ref{table:survey}, with most respondents being developers of an e-commerce product team, with a wide range of experience levels.

\begin{small}
\begin{table}
    \caption{Demographics of 16 Test Flow survey respondents.}
    \label{table:survey}
    \centering
    \begin{tabular}{ll|ll|ll}
        \hline
        \multicolumn{2}{|l|}{\textbf{Current Role}} & \multicolumn{2}{l|}{\textbf{Experience}} & \multicolumn{2}{l|}{\textbf{Curr. Prod.}}          \\ \hline
        \multicolumn{1}{|l|}{Developer}      & 11   & \multicolumn{1}{l|}{$<$ 1 year}      & 2    & \multicolumn{1}{l|}{E-commerce} & \multicolumn{1}{l|}{15} \\ \hline
        \multicolumn{1}{|l|}{ScrumMaster}   & 4    & \multicolumn{1}{l|}{1-3 years}             & 8    & \multicolumn{1}{l|}{Other}   & \multicolumn{1}{l|}{1}  \\ \hline
        \multicolumn{1}{|l|}{Intern}         & 1    & \multicolumn{1}{l|}{4-5 years}             & 4    &                              &                         \\ \cline{1-4}
        \textbf{}                            &      & \multicolumn{1}{l|}{$>$ 5 years}     & 2    &                              &                         \\ \cline{3-4}
    \end{tabular}
\end{table}
\end{small}

\begin{figure}
    \centering 
    \includegraphics[width=\linewidth]{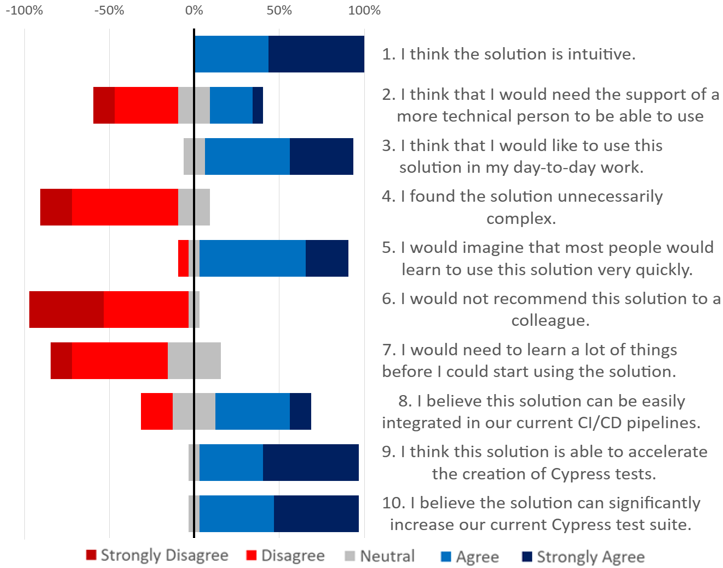} 
    \caption{Results of the Test Flow User Survey (16 Respondents).} 
    \label{fig:survey} 
\end{figure}

To evaluate Test Flow, ten statements were presented on a Likert scale from "Strongly Disagree" to "Strongly Agree." Respondents rated the solution's intuitiveness, accessibility, workflow integration, and potential to accelerate Cypress test script creation. Fig. \ref{fig:survey} shows the survey questions and results.

The statements were designed to ensure response validity by including both positively and negatively phrased questions. The overall feedback was highly positive, with consensus on 8 out of the 10 statements. Respondents found the solution intuitive, accessible, and believed it could be easily integrated into their workflows, enhancing the creation of Cypress tests.

However, statements 2 and 8 received mixed responses, leading to follow-up discussions with 11 respondents who opted for further engagement:
\begin{itemize}
    \itemsep 0em
    \item For statement 2, \textit{"I think that I would need the support of a more technical person to be able to use the solution"}, four respondents initially agreed, and one strongly agreed. 
    Upon follow-up, the four clarified that they misunderstood the question and found the solution user-friendly. The strongly agreeing respondent was anonymous and could not be contacted. 
    \item For statement 8, \textit{"I believe this solution can be easily integrated into our current CI/CD pipelines"}, three respondents disagreed, with only one available for follow-up. Feedback indicated the issue stemmed from specific CI/CD pipeline characteristics rather than the solution itself, suggesting that integration may require customization based on project needs.
\end{itemize}

During follow-up, participants also expressed concerns about data security when using Generative AI, as the use of company-hosted models was not emphasized in the presentation. Clarifying this point reassured participants.

Overall, Test Flow's validation provided valuable insights, highlighting its strengths and areas for improvement, including enhancing adaptability, refining integration support, and clearly communicating security measures.

\subsection{Answer to Research Questions}

Following the evaluation of both tools, we can address the research questions outlined in Section \ref{sec:intro}.

\vspace{4 pt}

\textbf{RQ1: To what extent can LLMs help in generating accurate and comprehensive acceptance test scenarios in natural language from user stories?}
    
In 65 usage instances, the users considered the ATs generated by AutoUAT helpful 95\% of the time. AutoUAT also received very positive feedback from 6 product owners.

We conclude that \textbf{AI-powered assistants like AutoUAT have a strong potential to help product owners and other stakeholders in creating high-quality acceptance test scenarios from user stories with less effort}.

\vspace{4 pt}

\textbf{RQ2: To what extent can LLMs help in generating accurate and complete executable test scripts from acceptance test scenarios in natural language?}

Test Flow generated accurate and complete executable test scripts from acceptance test scenarios in NL (with one test case per test scenario) 
60\% of the time, improved to 92\% with minor fixes (8\%) or additional contextual inputs (24\%). The generated scripts contained detailed comments in natural language to facilitate their (mandatory) review by developers responsible for test automation. Survey feedback was unanimously positive, emphasizing Test Flow's potential to speed up Cypress test creation and integrate smoothly into workflows.

We conclude that \textbf{AI-powered assistants like Test Flow have a strong potential to help developers in the translation of NL acceptance test scenarios to executable test scripts over the application UI with less effort}.

\vspace{4 pt}

However, further studies in industrial settings are needed to better assess the long-term productivity and quality gains.

\subsection{Threats to Validity}

Several factors may limit the generalizability of our findings:

\begin{itemize}
    \item \textbf{Context-Specific Results:} Evaluation was performed within a single company with strong quality processes. The approach may yield different outcomes in other environments or domains, especially where process or product standards are lacking (e.g., \textit{data-test-ids} for testability).

    \item \textbf{Sample Size Limitations:} The relatively small number of user stories tested with Test Flow and AutoUAT may affect the generalizability of results. A larger, more varied sample across different projects would enhance reliability.

    \item \textbf{Short Evaluation Period:} The two-month evaluation may miss long-term usability issues, evolving user behaviors, and learning effects.
        
    \item \textbf{Feedback Bias:} Out of 166 usage instances of AutoUAT, feedback was provided for only 65 instances, which may not represent all users. Positive experiences might be overrepresented, skewing acceptance rates.

\end{itemize}

These limitations indicate a need for further research and validation across diverse industrial settings to fully understand the approach's effectiveness and constraints.

\section{Related Work}\label{sec:related}

This section reviews existing literature and approaches related to our work, highlighting their results, limitations, and the unresolved challenges they present. By analysing these approaches, we aim to underscore the gaps our methodology addresses.

\subsection{From User Stories to NL Acceptance Test Scenarios}

The literature addressing automatic AT generation, specifically using LLMs within a BDD framework, is limited.  In fact, in \cite{wang2024software}, the authors identify several studies applying LLMs to unit and system testing but note that “there is still no research on the use of LLMs in integration testing and acceptance testing”, underscoring a research gap our work seeks to fill.

A notable example is a recent study \cite{10506519} that evaluates the capabilities of several LLMs (GPT-3.5, GPT-4, Llama-2-13B, and PaLM-
2) for generating syntactically correct Gherkin scenarios from user stories. In an experiment with 50 user stories extracted from various sources, both GPT models demonstrated superior performance, with syntax errors in only one out of 50 generated feature files. They also manually checked some generated acceptance tests for semantic validity.

In comparison, our approach uses the more recent GPT-4 Turbo model and includes a more comprehensive assessment of the generated Gherkin scenarios by tool users (typically product owners) in a real-world industrial setting, as well as initial feedback on potential performance benefits, thus providing empirical insights into the practical impact of LLMs for BDD. This industry-focused evaluation highlights the novelty and relevance of our approach.


\subsection{From Acceptance Test Scenarios to Executable Test Scripts}

We identified two recent studies \cite{leotta2024ai,leotta2024empirical} that examine productivity gains from using AI assistants and other acceleration technologies for generating test scripts compared to manual methods. Both studies start with test cases specified in Gherkin, which are then converted into executable tests.

In \cite{leotta2024ai}, the authors compare the effort to create Web end-to-end (E2E) test scripts based on natural language test descriptions using several approaches:

\begin{itemize}
    \item Manual - test scripts are crafted manually by the tester;
    \item ChatGPT Lite - the tester prompts ChatGPT to generate a test script based on the test description, copies the generated script to the IDE and manually refines it as needed, without further interaction with ChatGPT;
    \item ChatGPT Max - similar to Lite, but the tester may engage in a conversation with ChatGPT to refine the initial test script, before copying it to the IDE;
    \item Copilot - inside the IDE, the tester starts by providing the test description in natural language, which guides Github Copilot in the incremental generation of code suggestions, which are reviewed and edited as needed.
\end{itemize}

They used Gherkin for the description of the test scenarios in natural language, Java as the scripting language, JUnit as the test framework, and Selenium WebDriver as the browser automation library.

In an experiment involving 1 junior tester, 8 real-world Web applications, and 12 Gherkin test cases per application (crafted by one of the authors), they found that the Manual approach took the longest (median of 54 seconds per Gherkin line of code), while ChatGPT Max required the least time (38 seconds per line), with statistically significant productivity gains. ChatGPT Lite and Copilot had quite similar results (medians of 47 and 45 seconds per line, respectively).

In contrast, Test Flow's prompt includes the HTML code of the pages under test, drastically reducing the need for manual refinements to provide correct locators of UI elements, leading to potentially higher productivity gains.

In \cite{leotta2024empirical}, the authors compared the manual effort required for three Web test automation approaches:

\begin{itemize} 
    \item Programmable (PT) - manually implemented test cases; 
    \item Capture \& Replay (CRT) - captures manual interactions with a tool for later replay; 
    \item NLP-based (NLT) - describes test steps in a domain-specific language or restricted natural language, interpreted by a test automation tool, with special constructs for identifying the UI elements. 
\end{itemize}

The PT approach used Selenium WebDriver as the browser automation tool, Java as the programming language and JUnit as the test framework; the CRT approach used Selenium IDE to capture the user interactions, and NLT used an unspecified tool.

They conducted an experiment with 9 real-world Web applications with multiple releases, 3 junior testers/developers, and a set of Gherkin test cases written by one of the authors for each application (used as input specification for all approaches).

For the initial test suite development, PT required the most effort, CRT the least, with NLT in between. Over multiple releases, NLT had the lowest cumulative effort, except in one high-complexity case.
These results reflect CRT's low script reusability and NLT's limited flexibility for high-complexity applications. Thus, the authors conclude that NLT is well-suited for low to medium-complexity applications.

In contrast, Test Flow generates test scripts automatically from Gherkin test specifications with minimal user intervention. This is expected to reduce the initial test development effort significantly. In cases where automatic generation fails, users can resort to other approaches.

\subsection{Other Approaches}
We found a work \cite{zimmermann2023gui} that tries to use LLMs and ML to test Web applications through their GUI in a fully automatic way. However, the approach is closer to a monkey testing approach and not a specification-based one. They use GPT-4 and Selenium WebDriver for automatically exploring and testing Web applications in an interactive way without user intervention. 
In each step, they use Selenium WebDriver to capture the current state of the GUI and identify possible next actions on the GUI. Then, they prompt GPT-4 to select the next GUI action based on the following information: a generic test goal (such as maximizing the number of actions and states explored), the HTML code of the Web application, filtering out irrelevant formatting and behavioural information; the history of past actions in previous iterations; the possible next actions. The selected action is then performed in the AUT via Selenium WebDriver. This process is repeated until a specified number of interaction steps is reached.

In contrast, our approach is specification- and script-based, allowing for more comprehensive and repeatable specification coverage and error detection. Test Flow generates test scripts based on requirement specifications (user stories) and test specifications (Gherkin scenarios, including expected outputs), which, in turn, may be semi-automatically generated using AutoUAT. Overall, both approaches are complementary.

A few studies \cite{kamalakar2013automatically,storer2019behave} have explored generating executable tests from Gherkin test scenarios using NLP and automated reasoning techniques, but their targets were Java classes or Python functions: \cite{kamalakar2013automatically} achieved a 73\% success rate on a sample of 12 scenarios from tutorials, while \cite{storer2019behave} successfully translated 80\% of scenarios from 20 open-source projects used to develop the approach and 17\% from an additional 50 open-source projects. In contrast, our approach focuses on generating tests for Web applications through the GUI, posing distinct challenges, such as identifying target UI elements.

In \cite{wang2024xuat}, the authors introduce XUAT-Copilot, an LLM-powered tool for automating AT of mobile applications, specifically WeChat Pay. The tool takes a set of test cases composed of parameterized steps in natural language (similar to Gherkin) and corresponding test data as input. Three LLM-based agents—responsible for action planning, state checking, and parameter selection—collaborate to interpret these steps and translate them into UI actions on the app via an intermediate "skill" library, generating corresponding test scripts. 
In an experiment with a sample of passing test cases, XUAT-Copilot achieved 88.55\% success in test case translation (into passing test scripts), and 93.03\% in test step translation. 

In contrast, our approach targets Web applications, generating test scripts directly knowing the HTML of the pages under test, for a lighter solution. Although our approach already handles scenarios that involve navigation between multiple pages distinguishable by their URLs, web pages that dynamically change their structure and single-page applications may benefit from a more complex approach like XUAT-Copilot.


\section{Conclusions and Future Work}\label{sec:concl}

This section summarizes our key findings and suggests future research directions to enhance the use of LLMs in acceptance testing. 

\subsection{Conclusions}\label{ss:conclusions}

Our study demonstrates the potential of LLMs to improve the acceptance testing process, by automatically generating acceptance test scenarios and executable scripts from user stories. The solution, leveraging GPT-4 Turbo, was designed to streamline Gherkin scenario creation from user stories and Cypress script generation from these scenarios. Both tools received positive feedback, with AutoUAT rated highly by experienced product owners and Test Flow noted for its usability and effectiveness in automating test creation.

AutoUAT achieved a 95\% acceptance rate in AT generation, covering both explicit requirements and additional scenarios. Test Flow generated test scripts with 60\% initial accuracy, rising to 92\% with minor refinements, thus reducing developer workload in creating and verifying test scripts. 

While effective, integrating LLMs for AT generation has challenges, such as ensuring detailed contextual inputs (namely, product and UI descriptions), managing potential errors, and addressing cost and concerns. Despite these challenges, our results suggest that LLMs can significantly improve the acceptance testing process, with time and cost savings.

\subsection{Future Work}\label{ss:future_work}

This study suggests several directions for future research in the context of AI-assisted acceptance testing: training models on domain-specific datasets to enhance AT accuracy, integrating AT generation with issue management tools, exploring the impact of detailed user stories on test quality, optimizing costs by reducing input token requirements, conducting broader studies across projects and industries, and conducting long-term studies to better assess productivity and quality benefits.

\appendix

\subsection{Structure of AutoUAT's prompt}\label{sec:AutoUATprompt}

\begin{lstlisting}[breaklines=true, breakindent=0pt, basicstyle=\ttfamily\small, xleftmargin=0pt, numbers=none]
System:
You are an expert assistant specializing in creating User Acceptance Tests using Gherkin language. Your behavior should be that of a meticulous and detail-oriented professional, dedicated to producing clear, comprehensive, and precise User Acceptance Tests. Your objective is to generate high-quality  User Acceptance Tests based on the provided  User Story titles and descriptions, ensuring no ambiguities. The tests should be thorough and follow the Gherkin syntax accurately.

Context:
The User Acceptance Tests you generate will be used by a development team to validate that the functionality of the application meets the specified requirements. It is crucial that the tests cover various scenarios, including edge cases, to ensure robust validation. The User Stories provided will include a title and a description, and your task is to translate these into Gherkin language tests.

{% for item in chat_history %}
user:
Here is a User Story title and description. Generate User Acceptance Tests in Gherkin language for this User Story.

User Story Title: {{item.inputs.title}}
User Story Description: {{item.inputs.description}}

Here are the User Acceptance Tests in Gherkin language for the given User Story:
assistant:
{{item.outputs.answer}}
{% endfor %}
\end{lstlisting}

\subsection{Structure of Test Flow's System Prompt} \label{sec:TestFlowPrompt}

\begin{lstlisting}[breaklines=true, breakindent=0pt, basicstyle=\ttfamily\small,  xleftmargin=0pt, numbers=none]
(...Product context information omitted for confidentiality reasons...)

You will receive a user story, Gherkin scenarios and HTML of the pages to test. 
You are responsible for writing the Cypress tests for the Gherkin scenarios.

Keep in mind the following best practices:
- You generate the test to be as complete as possible for the scenario.
- You use the data-test-id to locate the element if you need to interact with it.
- Keep tests independent, so they can run in any order.
- Use Cypress built-in assertions.

You already have some Cypress commands and  variables to use in your tests:

```typescript
(...omitted...)
```

Some notes:
- You output the Cypress code only and inside a markdown code block.
- If any additional information is needed, put it in a comment inside the code block.
- Pay attention to the html provided when writing the tests so that you can use the correct data-testid and Cypress commands to interact with the elements.
- When using text to assert the content of an element, pay attention to the language of the page (...).
- Ensure your code includes comments that guide through the steps for code understanding and accessibility.
\end{lstlisting}

\subsection{Example of User Story Inputted to AutoUAT}\label{sec:example}

\textbf{\textit{Title:}} Alphabet User Sign-Up

\textbf{\textit{Description:}}
\begin{itemize}
    \item \textbf{As an} Alphabet App user,\\
    \textbf{I want} to access my profile,\\
    \textbf{So that} I can allow for the collection of Analytics.
\end{itemize}
When the user is in the Profile \& Settings Main Page, in the Legal Information section the user can activate or deactivate a toggle to allow the collection of data (analytics). This will communicate with the Backend to guarantee the activation for analytics tracking.

\textbf{Actions \& Error Handling}:
\begin{itemize}
    \item The user clicks on the toggle. It moves to the right and becomes green to activate.
    \item The user clicks on the toggle. It moves to the left and becomes grey to deactivate.
\end{itemize}

\subsection{Example of Generated Acceptance Test Scenarios by AutoUAT}\label{sec:uatexample}

\begin{lstlisting}[caption={Example of Generated Acceptance Test Scenarios by AutoUAT},captionpos=b,language=Gherkin]
Feature: Legal Information - Usage Data Tracking

Scenario: User activates the collection of analytics data
Given the user is on the Profile & Settings Main Page
And the user is in the Legal Information section
When the user clicks on the data collection toggle
Then the toggle should move to the right
And the toggle should turn green
And the backend should be notified to activate analytics tracking

Scenario: User deactivates the collection of analytics data
Given the user is on the Profile & Settings Main Page
And the user is in the Legal Information section
When the user clicks on the data collection toggle
Then the toggle should move to the left
And the toggle should turn grey
And the backend should be notified to deactivate analytics tracking

Scenario: User tries to activate the collection of analytics data but the backend fails
Given the user is on the Profile & Settings Main Page 
And the user is in the Legal Information section
When the user clicks on the data collection toggle
And the backend fails to activate analytics tracking
Then the toggle should not move to the right
And the toggle should not turn green
And an error message should be displayed
    
Scenario: User tries to deactivate the collection of analytics data but the backend fails
Given the user is on the Profile & Settings Main Page 
And the user is in the Legal Information section 
When the user clicks on the data collection toggle 
And the backend fails to deactivate analytics tracking 
Then the toggle should not move to the left
And the toggle should not turn grey
And an error message should be displayed 
\end{itemize}
\end{lstlisting}
\subsection{Example of Test Flow's Inputs}

\begin{lstlisting}[caption={Example of Test Flow's Inputs: The User Story},captionpos=b,language=US]
As a customer, 
I want to see detailed information of phyiscal products on the detail page, 
So that I can see specific details of the product, like its material.
\end{lstlisting}

\begin{lstlisting}[caption={Example of Test Flow's Inputs: The Gherkin Test Cases},captionpos=b,language=Gherkin]
Feature: Accordion with texts on detail page  

Scenario: Display first section unfolded when customer opens the page 
Given the customer is on the product detail page 
When the page is loaded 
Then the first section of the accordion should be displayed unfolded  

Scenario: Unfold and collapse sections via arrow 
Given the customer is on the product detail page 
When the customer clicks on the arrow next to a section 
Then that section should toggle between unfolded and collapsed states  
\end{lstlisting}
\subsection{Example of Test Flow's Generated Cypress Test Script}

\begin{lstlisting}[caption={Example of Test Flow's Generated Cypress Test Script},numbers=none,xleftmargin=0pt, captionpos=b,language=JavaScript]
describe('Accordion with texts on detail page', () => {
  const productDetailPageUrl = (*\censor{$'/bmw_DE_de-DE/shop/ls/dp/physical-goods/900653'$};*)
  beforeEach(() => {
    cy.setTestCookies();
    cy.login(Cypress.env('users').DE);
    cy.visit(productDetailPageUrl);
    cy.setTestDealer();
    cy.reload();
    cy.wait(2000);
  });

  it('Display first section unfolded when customer opens the page', () => {
    // Check if the first section of the accordion is displayed unfolded
    cy.get('[data-testid="accordion-item-0"]').within(() => {
      cy.get('h2').should('have.text', 'Produktdetails');
      cy.get('.accordion-item-children').should('be.visible');
    });
  });

  it('Unfold and collapse sections via arrow', () => {
    // Click on the second section to unfold it
    cy.get('[data-testid="accordion-item-1"] h2').click();
    cy.get('[data-testid="accordion-item-1"] .accordion-item-children').
        .should('be.visible');
    // Click again to collapse it
    cy.get('[data-testid="accordion-item-1"] h2').click();
    cy.get('[data-testid="accordion-item-1"] .accordion-item-children')
        .should('not.be.visible');
  });
});
\end{lstlisting}

\bibliographystyle{IEEEtran}
\bibliography{cas-refs}
\end{document}